\documentclass[a4paper,11pt]{article}
\usepackage{amsmath}
\usepackage{amssymb}
\usepackage{graphicx}
\usepackage{cite}
\usepackage[T1]{fontenc}
\usepackage[a4paper,left=1.0in, right=1.0in,top=1.1in, bottom=1.2in]{geometry}

\numberwithin{equation}{section}

\newcommand{\bit}{\begin{itemize}}
\newcommand{\eit}{\end{itemize}}
\newcommand{\be}{\begin{equation}}
\newcommand{\ee}{\end{equation}}
\newcommand{\bea}{\begin{eqnarray}}
\newcommand{\eea}{\end{eqnarray}}
\newcommand{\bsp}{\begin{split}}
\newcommand{\esp}{\end{split}}

\title{ 
    \vskip 2cm
    Gluon saturation in dijet production in p-Pb collisions\\
    at Large Hadron Collider
}

\author{
  Krzysztof Kutak \\
  {\it Instytut Fizyki J\k{a}drowej im. Henryka Niewodnicza\'nskiego}\\
  {\it Radzikowskiego 152, 31-342 Krak\'ow, Poland}\\ \\
  Sebastian Sapeta \\
  {\it Institute for Particle Physics Phenomenology, Durham University,}\\
  {\it South Rd, Durham DH1 3LE, UK}
}
\date{}

\begin{document}
\maketitle
\vspace{-25em}
\begin{flushright}
  IPPP/12/35\\
  DCPT/12/70\\
  IFJPAN-IV-2012-1
\end{flushright}
\vspace{20em}

\begin{abstract}
We study saturation effects in the production of dijets in p-p and p-Pb
collisions using the framework of {\it high energy factorization}. We focus on
central-forward jet configurations, which allow for probing gluon density at low
longitudinal momentum fraction.  
We find significant suppression of the central-forward jet decorrelations in
p-Pb compared to p-p, which we attribute to saturation of gluon density in
nuclei.
\end{abstract}

\section{Introduction}

Physics in the forward region at the Large Hadron Collider (LHC) is a very
interesting and exciting field since it involves interplay of the kinematical
scales, like for example transverse momenta of jets, with scales generated by
the QCD dynamics, like the saturation scale $Q_s$ \cite{Gribov:1984tu}.
The latter scale characterizes formation of dense system of partons and there is
a growing evidence that the phenomenon of saturation of gluons indeed occurs
\cite{Stasto:2000er,Albacete:2010pg,Dumitru:2010iy, Albacete:2012rx}. 
To further advance studies of saturation and other possible effects occurring
at high partonic density, the LHC is going to collide p-Pb this year.
This will allow, in particular,  for the study of the onset of saturation as a
function of variables related to the transverse momentum $p_t$ of jets.
In addition, the understanding of the interplay of scales in the jet production
in p-p and p-Pb will permit to constrain the unintegrated parton densities in
the large phase space available for partons, i.e.  $10^{-6}\!<\!x\!<\!0.1$,
$5\,\text{GeV}\!<\!k_t\!<\!150\,\text{GeV}$, where $x$ is the longitudinal
momentum fraction of the hadron carried by a parton, while $k_t$ is the
component of its momentum transverse to the collision axis.
In particular, the study of exclusive final states, like jets, allows for
determination of the unintegrated gluon density in the range of large momenta.
 
There are formalisms that allow one to study dense systems \cite{Gelis:2010nm}
or systems with hard momentum scale involved \cite{Sjostrand:2006za}.  However,
the formalism which accounts for both the high energy scale and the hard
momentum scale $p_t$ is provided only by the \emph{high energy factorization}
\cite{Catani:1990eg}.  
In this framework, one of the elements that enter the factorization formula is
the unintegrated gluon density.  Depending on approximation, it satisfies the
BFKL, BK, or CCFM evolution
equations~\cite{Kuraev:1977fs,Balitsky:1978ic,Balitsky:1995ub,Kovchegov:1999yj,Kovchegov:1999ua,Ciafaloni:1987ur,Catani:1989sg,Catani:1989yc}.
The BFKL and BK equations are known already at NLO and sum up emissions of
gluons with strong ordering in the longitudinal momentum fractions of
subsequently emitted gluons.  The important issue for phenomenological
applications, in particular for exclusive final states at large~$p_t$, is to
perform resummations of most relevant higher order corrections to the
evolution kernel of the BFKL equation.  This is because only then, the solution
of the equation for the unintegrated gluon density is physically relevant and
well defined \cite{Salam:1998tj,Ciafaloni:1998iv,Ciafaloni:2003kd}.

Another issue is that since the BFKL or CCFM equation are linear they
predict strong rise of gluon density at small values of gluon's $k_t$ which
leads to conflict with unitarity bounds. 
Effects of higher orders, although suppress the growth of gluon density,
preserve its power like behavior as a function of $k_t$ at low $k_t$ values. To
restore unitarity one supplements the BFKL equation with a nonlinear term which
accounts for fusions of gluons. These unitarity corrections, which are taken
into account in the BK or JIMWLK equations, give rise to an emergent semi-hard
scale, called the saturation scale $Q_s(x)$, at which the gluon density has a
maximum and which therefore defines the most probable momentum of gluon
\cite{Kutak:2009zk,Kovchegov:2010pw}.  
 
An interesting process in which both saturation and production at high
$p_t$ can be studied, is the process of dijet production where the jets are
separated by large rapidity \cite{Marquet:2003dm,d'Enterria:2006nb,Iancu:2006uc,Andersen:2003an}.
More specifically, we shall focus on the case in which one jet is in the central
while the other in the forward rapidity region. 
Such a final state probes parton density of one of the protons at low
longitudinal momentum fraction $x$ while the other at large longitudinal
momentum fraction. 
The latter proton can be described by the well known collinear parton
distribution functions and therefore such process is perfectly suited to study
properties of the unintegrated gluon density at low $x$ and especially its
saturation.
 
In this paper, we discuss production of dijets in the p-p and p-Pb collisions.
The former serves as a benchmark, i.e. first we fit the unintegrated gluon
density to the $F_2$ data \cite{Aaron:2009aa} and then apply it to calculate
observables in p-p characterizing the dijet system like angular correlations of
produced jets, $p_t$ spectra of forward and central jets and their rapidity
distributions.
In the next step we compute predictions for rapidity distributions and angular
decorrelations of central-forward jets produced in p-Pb collision. For both
observables we see significant suppression of the cross section due to
saturation effects in the nucleus.

The paper is organized as follows. In section \ref{sec:hef} we introduce the
{\it high energy factorization} framework and define the observables we want to
use as a tool to study saturation effects. In section \ref{sec:fit} we introduce
the unified BK/DGLAP equation \cite{Kutak:2003bd} for the unintegrated
gluon density and present results of fits of the unintegrated gluon to the
combined HERA data.  In section \ref{sec:results-pp}  we apply the high energy
factorization framework together with our fitted unintegrated gluon to calculate
observables for central-forward dijet system. In the section
\ref{sec:results-p-Pb} we calculate rapidity distribution and angular
decorrelations of central-forward jets produced in p-Pb collision. We conclude
our studies in section \ref{sec:conclusions}.

\section{Dijet production in high energy factorization approach}
\label{sec:hef}

\begin{figure}[t] \centering
  \includegraphics[width=0.45\textwidth]{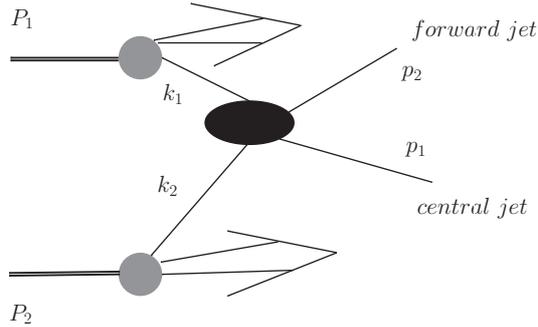}
  \caption{Jet production in the forward region in hadron-hadron collisions.
  } 
  \label{fig:jet_production} 
\end{figure}

The main goal of this paper is to provide predictions for azimuthal
decorrelations of jets produced in p-p and p-Pb collisions.  Consider the
process of the production of a dijet system in the collision of two hadrons  
\be
  A+B \to J_1 + J_2+X\,.
\ee
The leading order contribution comes from the $2\to 2$ partonic process
\be
  a (k_1) + b (k_2) \to c (p_1)+ d (p_2)\,.
\ee
In this study we focus on the asymmetric configuration with one jet produced in
the forward and the other in the central rapidity region as illustrated in Fig.~\ref{fig:jet_production}. 
The fractions of the longitudinal momenta of the initial state partons are
related to the transverse momenta and rapidities of the final state partons by
\be
  x_1 = \frac{1}{\sqrt{S}} \left(p_{t1} e^{y_1} + p_{t2} e^{y_2}\right)\,,
  \qquad
  x_2 = \frac{1}{\sqrt{S}} \left(p_{t1} e^{-y_1} + p_{t2} e^{-y_2}\right)\,,
  \label{eq:x1x2}
\ee
where $S$ is the squared energy in the center of mass system of the incoming
hadrons.
Hence, our central-forward configuration corresponds to one of the $x_i$s
being small and the other one large.

The first study of such configurations in the p-p collisions was performed in
\cite{Deak:2010gk,Deak:2009ae,Deak:2011ga} using the CASCADE Monte Carlo generator \cite{Jung:2010si}. Assuming, without loss of 
generality, that $x_1\simeq 1$ and
$x_2 \ll 1$ we have
\bea
  \label{eq:k1-approx}
  k_{1}^{\mu} & = & x_1 P_1^{\mu}\,,
  \\
  \label{eq:k2-approx}
  k_{2}^{\mu} & = & x_2 P_2^{\mu} + k_t^{\mu}\,,
\eea
where we used the Sudakov decomposition of the initial partons' 4-momenta.  Here
$P_{1,2}^{\mu}$ are the 4-momenta of incoming hadrons, which, in the center of
mass frame, take the form $P_{1,2}^{\mu} = \sqrt{\frac{s}{2}}\left(1, 0, 0,
\pm1\right)$ and $P_{1}\cdot P_{2} = \frac12 S$. The momentum of the off-shell
parton satisfies $k_2^2=k_t^2\equiv -k^2$, 
where $k\equiv k_t \equiv |\bf{k}|$.
This leads to the following form of the cross section
\be
  \frac{d\sigma}{dy_1dy_2d^2p_{1t}d^2p_{2t}} 
  =
  \sum_{a,c,d} 
  \frac{1}{16\pi^3 (x_1x_2 S)^2}
  {\cal M}_{ag\to cd}
  x_1 f_{a/A}(x_1,\mu^2)\, 
  \phi_{g/B}(x_2,k^2,\mu^2)\frac{1}{1+\delta_{cd}}\,,
  \label{eq:cs-fac}
\ee
and
\be
  k^2 = p_{t1}^2 + p_{t2}^2 + 2p_{t1}p_{t2} \cos\Delta\phi\,,
\ee
where $\Delta\phi=\phi_1-\phi_2$ is the azimuthal distance between the outgoing
partons and ${\cal M}_{ag\to cd}$ is the matrix element for the $2\to 2$
process with one off-shell initial state gluon and three on-shell partons,
$a,c,d$, which
can be either quarks or gluons \cite{Deak:2009xt}. 
The following partonic sub-processes contribute to the production of our dijet
system
\be
  qg  \to  qg\,,
  \qquad \qquad 
  gg  \to  q\bar q\,,
  \qquad \qquad 
  gg  \to  gg\,.
\ee
On the side of the off-shell gluon in Eq.~(\ref{eq:cs-fac}), we have the
unintegrated gluon density $\phi_{g/B}(x_2,k^2\mu^2)$, which depends on the
longitudinal momentum fraction $x_2$, on the transverse momentum of the
off-shell gluon, and in general as well as on hard scale $\mu$.  The hard scale
dependence introduces DGLAP-like ordering effects in the high energy
factorization framework and makes it applicable in studies of exclusive final
states. In our calculations, however, we follow the KMS \cite{Kwiecinski:1997ee} scheme to introduce
corrections to the gluon density which make it applicable to the studies of jet
physics. Because of this we shall skip the argument $\mu$ in the expressions for
the unintegrated gluon density below.
 
On the side of the on-shell parton, which is probed at high values of the
longitudinal momentum fraction $x_1$, it is legitimate to use the collinear
parton density $f_{a/A}(x_1,\mu^2)$. 

The above result depends only on the difference of the azimuthal angles
$\Delta\phi$, so one can change variables and integrate out one of angles
$\phi_i$. This leads to
\be
  \frac{d\sigma}{dy_1dy_2dp_{1t}dp_{2t}d\Delta\phi} 
  =
  \sum_{a,c,d} 
  \frac{p_{t1}p_{t2}}{8\pi^2 (x_1x_2 S)^2}
  {\cal M}_{ag\to cd}
  x_1 f_{a/A}(x_1,\mu^2)\,
  \phi_{g/B}(x_2,k^2)\frac{1}{1+\delta_{cd}}\,,
  \label{eq:cs-main}
\ee
with  $k^2 = p_{t1}^2 + p_{t2}^2 + 2p_{t1}p_{t2} \cos\Delta\phi$.

\section{Unintegrated gluon density from the unified BK/DGLAP framework fitted
to combined HERA data}
\label{sec:fit}

The formulation of the NLO BFKL
equation~\cite{Fadin:1996zv,Fadin:1997hr,Fadin:1998py} has been known already
for some time. Also the NLO BK equation has been derived \cite{Balitsky:2008zza}
but, because of its complicated structure, only solutions of some approximate
forms of the BK equations are known (see \cite{Triantafyllopoulos:2002nz,
Peschanski:2006bm, Enberg:2006aq,Albacete:2010sy}).
The basic formulation of the NLO BFKL equation is unstable (due to non-positive
definite kernel) and in order to stabilize it one needs to resume a subset of
higher order corrections \cite{Kwiecinski:1997ee,Salam:1998tj,Ciafaloni:1998iv}.
In our study, we will use the approach to this problem formulated in
\cite{Kwiecinski:1997ee} in which large part of the higher order corrections is
provided by the consistency constraint on emissions of real gluons.
The other important corrections are coming from running of the coupling
constant and the nonsingular pieces of the DGLAP splitting functions.  Other
approaches were discusses in \cite{Kowalski:2010ue,Vera:2005jt,Gotsman:2002yy}. 

\begin{figure}[t] \centering
  \includegraphics[width=0.35\textwidth]{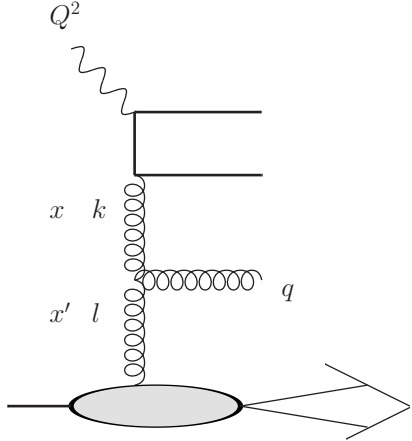}
  \caption{Diagrammatic representation of the $F_2$ structure function in {\it high energy factorization}.
  } 
  \label{fig:kinematyka} 
\end{figure}

\begin{figure}[t] 
  \centering
  \includegraphics[height=0.62\textheight]{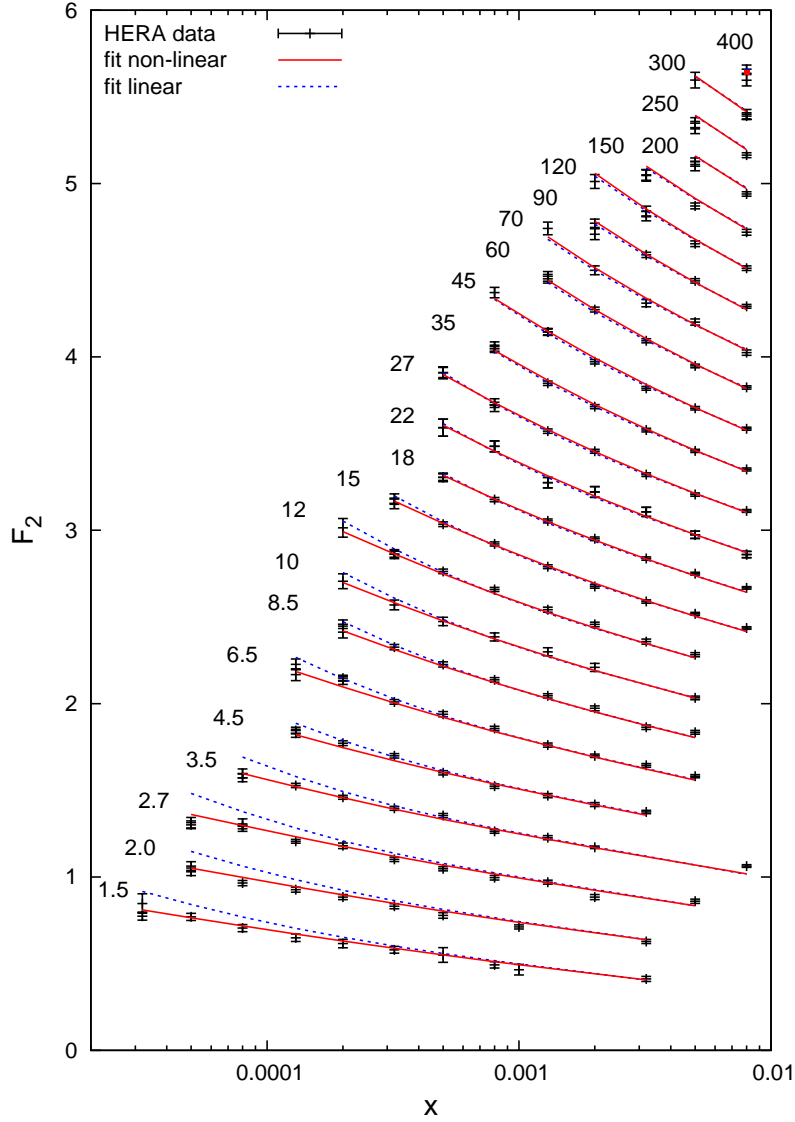}
  \caption{
  The proton structure function $F_2(x,Q^2)$ from the fit of our framework, in
  its linear and nonlinear variant, to the combined data from HERA
  \cite{Aaron:2009aa} as a function of $x$ for the $Q^2$ range from 1.5 to 400 
  GeV$^2$ (with the vertical offsets of 0.2).
  } 
  \label{fig:f2plot} 
\end{figure}

\begin{figure}[t] 
  \centering
  \includegraphics[width=0.32\textwidth]{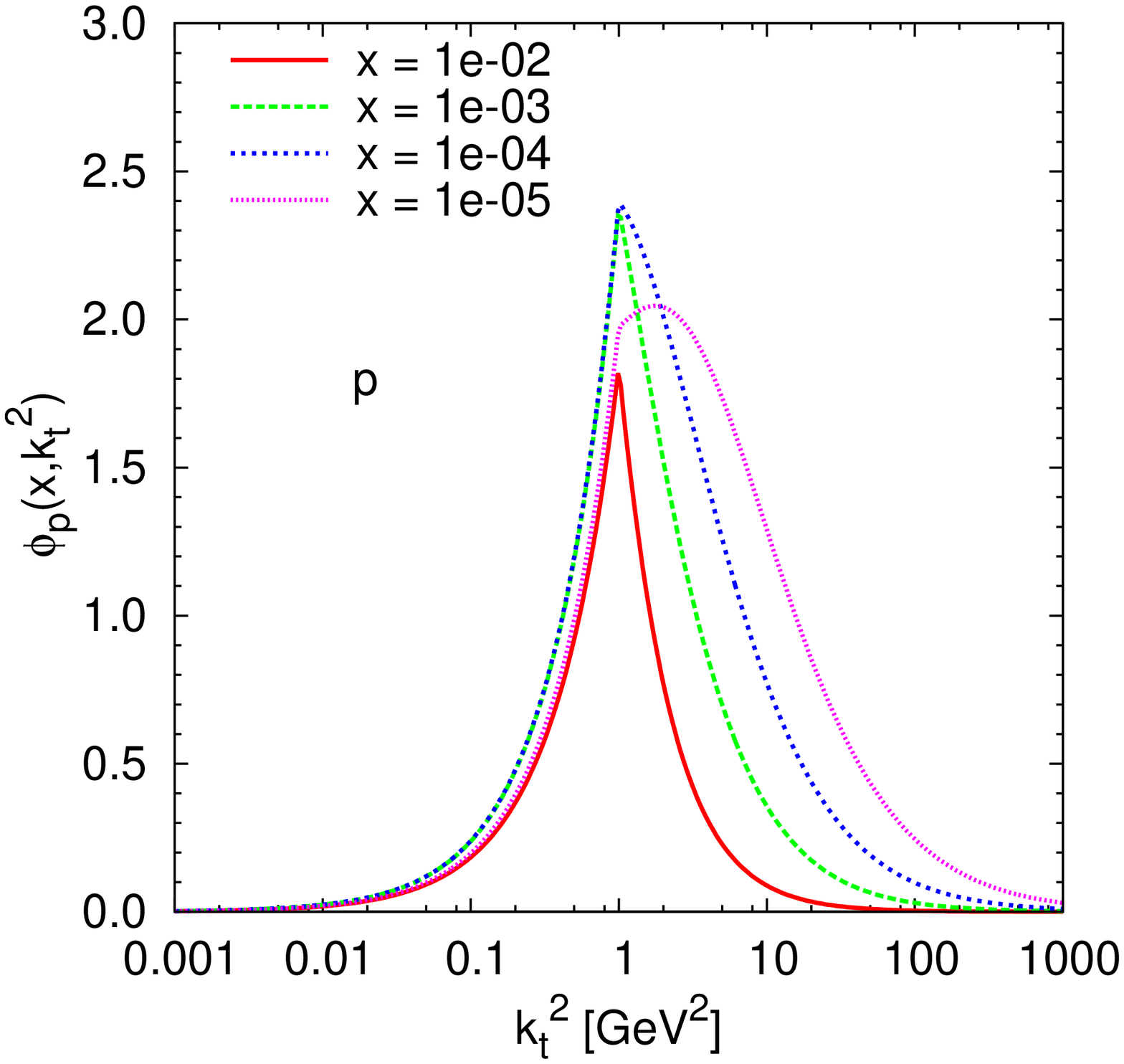}
  \includegraphics[width=0.33\textwidth]{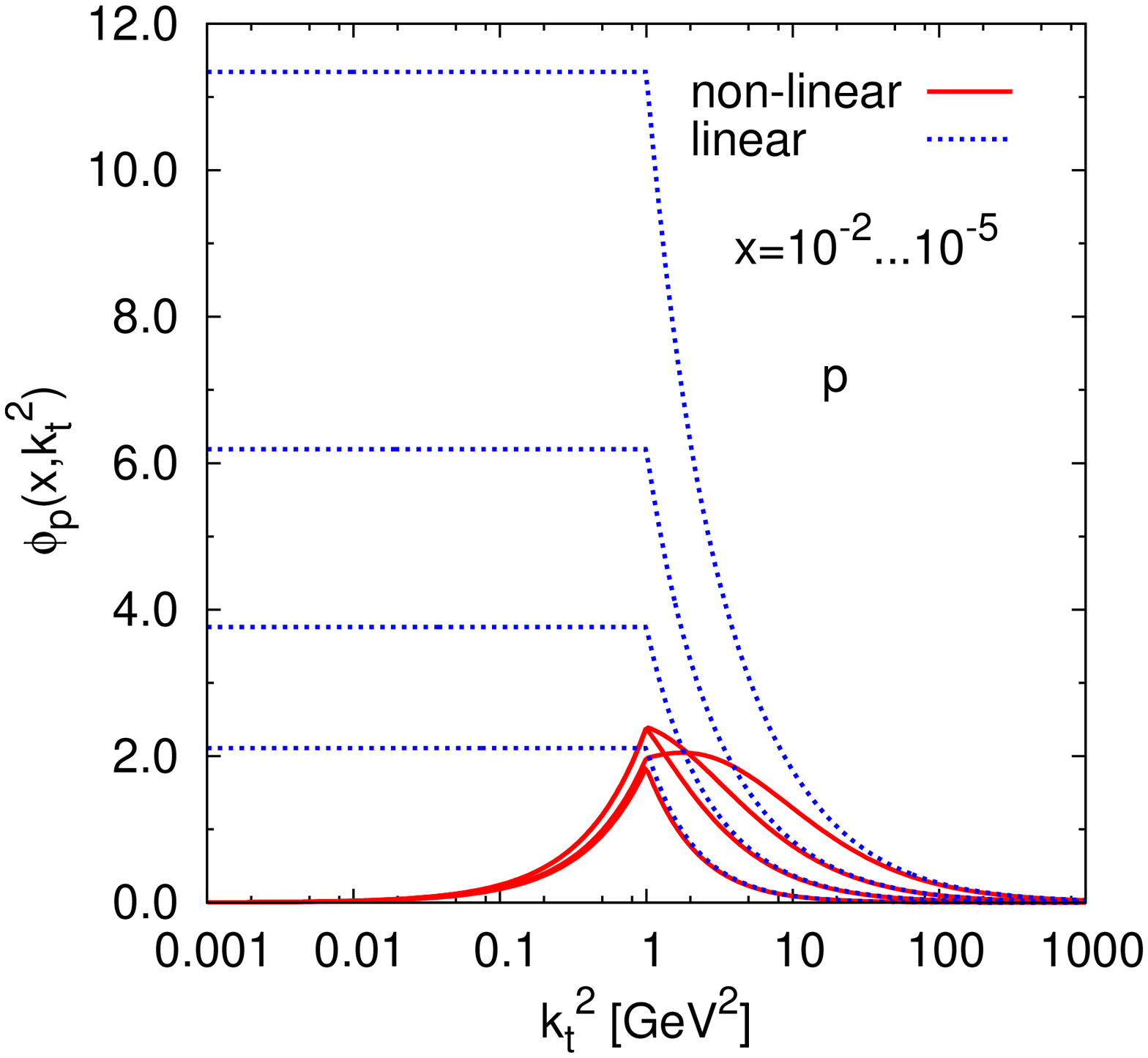}
  \includegraphics[width=0.32\textwidth]{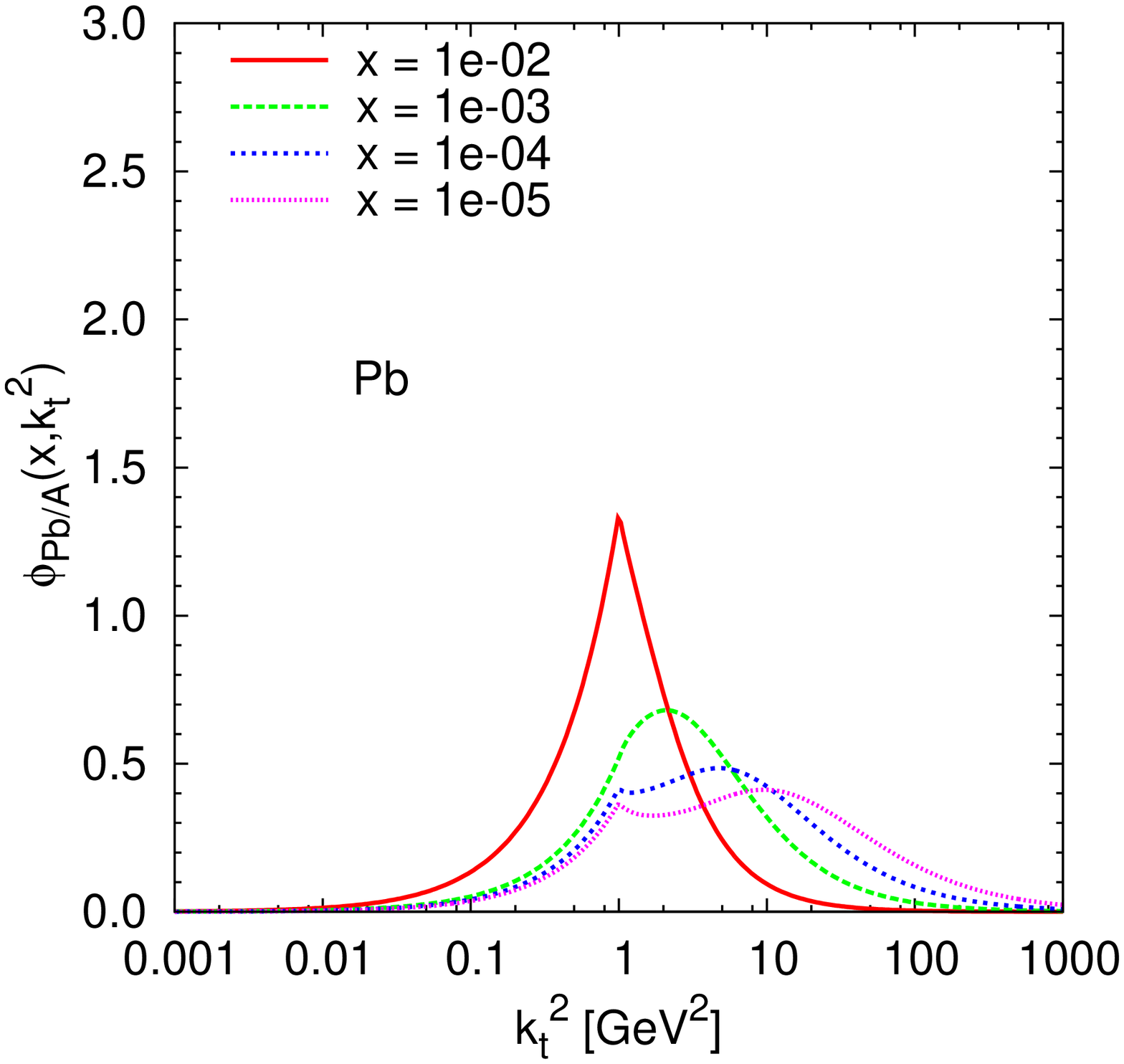}
  \caption{ 
  Left: Unintegrated gluon density in the proton from the solution of
  Eq.~(\ref{eq:fkovres}) with free parameters fitted to $F_2$ data. 
  Middle: Comparison of the gluon density from the left plot with the
  unintegrated gluon from the fit using only the linear part of
  Eq.~(\ref{eq:fkovres}).
  Right: Unintegrated gluon density in the Pb nucleus obtained as for the
  proton case but with the nonlinear term in Eq.~(\ref{eq:fkovres}) enhanced 
  by $A^{1/3}$ with $A=207$.
  }
  \label{fig:ugluon} 
\end{figure}

The corresponding equation for the unintegrated gluon density reads \cite{Kutak:2003bd,Kutak:2004ym} 
\begin{multline} 
\phi_p(x,k^2) \; = \; \phi^{(0)}_p(x,k^2) \\
+\,\frac{\alpha_s(k^2) N_c}{\pi}\int_x^1 \frac{dz}{z} \int_{k_0^2}^{\infty}
\frac{dl^2}{l^2} \,   \bigg\{ \, \frac{l^2\phi_p(\frac{x}{z},l^2) \, \theta(\frac{k^2}{z}-l^2)\,   -\,
k^2\phi_p(\frac{x}{z},k^2)}{|l^2-k^2|}   +\,
\frac{k^2\phi_p(\frac{x}{z},k^2)}{|4l^4+k^4|^{\frac{1}{2}}} \,
\bigg\} \\  + \, \frac{\alpha_s (k^2)}{2\pi k^2} \int_x^1 dz \,
\Bigg[\left(P_{gg}(z)-\frac{2N_c}{z}\right) \int^{k^2}_{k_0^2} d l^{
2}\,\phi_p\left(\frac{x}{z},l^2\right)+zP_{gq}(z)\Sigma\left(\frac{x}{z},k^2\right)\Bigg] \\
-\frac{2\alpha_s^2(k^2)}{R^2}\left[\left(\int_{k^2}^{\infty}\frac{dl^2}{l^2}\phi_p(x,l^2)\right)^2
+
\phi_p(x,k^2)\int_{k^2}^{\infty}\frac{dl^2}{l^2}\ln\left(\frac{l^2}{k^2}\right)\phi_p(x,l^2)\right]\,,
\label{eq:fkovres} 
\end{multline} 
where $z=x/x'$ (see Fig. \ref{fig:kinematyka} for explanation of the variables).
For convenience, we omit the~$g$ subscript in the unintegrated
gluon density symbol and keep only the subscript denoting the hadron. 
The theta function in Eq.~(\ref{eq:fkovres}) introduces the kinematical
constraint and the nonlinear term, which supplies unitarity corrections, is
given by the triple pomeron vertex \cite{Bartels:1994jj}. The two terms in the third line in Eq. (\ref{eq:fkovres})
correspond to the DGLAP effects generated by that part of the splitting function
$P_{gg}(z)$ which is nonsingular in the limit $z\rightarrow 0$ and by the quarks respectively, with $\Sigma(x, k^2)$ corresponding
to the singlet quark distributions. The   At LO in $\ln1/x$
this equation reduces to the BK equation, after performing Fourier transform to
the coordinate space
\cite{Kovchegov:1999ua,Bartels:1994jj,Kutak:2003bd,Bartels:2007dm}. For numerical method to solve Eq.~(\ref{eq:fkovres}) we refer the reader to \cite{Kutak:2003bd}.

The strength of the nonlinear term in Eq.~(\ref{eq:fkovres}) is controlled by
the parameter $R$ which has an interpretation of the proton radius and it comes
from integration of the gluon density over the impact parameter $b$. 
In our framework, we assume the uniform distributions of gluons in the nucleon
therefore our gluon density is proportional to $\Theta(R-b)$
\cite{Kutak:2003bd}. 

The input gluon distribution $\phi^{(0)}_p(x,k^2)$ is is given by 
\be
 \phi^{(0)}_p(x,k^2)=\frac{\alpha_S(k^2)}{2\pi k^2}\int_x^1
 dzP_{gg}(z)\frac{x}{z}g\left(\frac{x}{z},k_0^2\right)\,,
 \label{eq:initial-cond} 
\ee 
where $xg(x,k_0^2)$ is the integrated gluon distribution at the initial scale,
which we set to $k_0^2=1\, \text{GeV}^2$, and we take the following
parametrization
\be
 xg(x)=N(1-x)^{\beta} (1-D x)\,,
 \label{eq:input-param} 
\ee 
which is similar to what was used in ~\cite{Bacchetta:2010hh}. For the strong
coupling, we take the one-loop result with $\Lambda_{\text{QCD}}$ set to 350
MeV.

The evolution equation~(\ref{eq:fkovres}) is used to determine the
unintegrated gluon above the initial momentum scale that is for $k^2>1\,
\text{GeV}^2$. In the region $k^2<1\, \text{GeV}^2$,
the gluon density $\phi_p(x,k^2)$ is constrained by the condition that it should
match the evolved unintegrated gluon density at $k^2=k^2_0$ . 
We choose to parametrize the unintegrated gluon in this region by
\be
\phi_p(x,k^2) = 
k^2 \phi_p(x,k_0^2=1\,\text{GeV}^2)\qquad \text{for} \quad k^2 < 1\, \text{GeV}^2\,,
\label{eq:ug-belowkt1}
\ee
which is motivated by the shape of the gluon density obtained from solution of
the LO BK equation in the saturated regime \cite{Sergey:2008wk}.

The unintegrated gluon density from
Eqs.(\ref{eq:fkovres})--(\ref{eq:input-param}) convoluted with impact factors
taken from \cite{Kwiecinski:1997ee} allows one to compute the structure function
$F_2(x,Q^2)$, which in turn can be used to fit the free parameters of the model.
We performed such a fit to the combined HERA data~\cite{Aaron:2009aa} in the
kinematical range of $x<0.01$ and the full range of $Q^2$. As shown in
Fig.~\ref{fig:f2plot} (red solid line), we obtain a very good description of
data, which corresponds to $\chi^2/\text{ndof} = 1.73$ and the following values
of the parameters: $N=0.994$, $\beta= 18.6$, $D=-82.1$ and $R=2.40\,
\text{GeV}^{-1}$.
 
In Fig.~\ref{fig:ugluon}~(left) we show the unintegrated gluon density,
corresponding to the above fit, as a function of the transverse momentum of the
gluon for a range of $x$ values. The gluon from the evolution is sewed at
$k^2=1\, \text{GeV}^2$ with the parametrization (\ref{eq:ug-belowkt1}). 
The sharp peek corresponds to the point in $k$ where the matching was done. We
see however that as one goes to lower $x$ values, perturbatively generated
maximum starts to emerge. That is a signal of the presence of saturation scale
defining the most probable momentum of the gluon. The results for estimated
value of the saturation scale are in an agreement with
\cite{Marquet:2005vp,Sergey:2008wk,Albacete:2010sy}

Our main focus in this study is on the framework with saturation of gluon
density described above which is based on the nonlinear evolution equation
(\ref{eq:fkovres}). It is however interesting to compare our results to
the case in which the gluon is determined from the framework without saturation.
This is naturally provided by the linear version of Eq.~(\ref{eq:fkovres}) that 
corresponds to dropping the last term on the right hand side of that equation,
which now becomes independent of $R$.
We performed an analogous fit to the one described above but taking the
linearized version of Eq.~(\ref{eq:fkovres}) and restricting $Q^2$ to the values
above $4.5\, \text{GeV}^2$ to stay outside of the region where the saturation
effects may be important.
The fit parameters at the
minimal value of $\chi^2=1.51$ are: $N=0.004$, $\beta=26.7$ and $D=-51102$. The
corresponding results are shown in Fig.~\ref{fig:f2plot}~(blue dashed line) for the whole range of
$Q^2$ including the bins below $4.5\, \text{GeV}^2$, which were not
used in the fit. We see that the linear gluon gives too strong rise of $F_2$
with $x$ especially at low values of $Q^2$. This remains true even if we
fit the linear version of Eq.~(\ref{eq:fkovres}) to the full range of $Q^2$.
The best value of $\chi^2$ we were able to achieve in this case was 3.86. 
We therefore conclude that some mechanism damping the gluon density at low $x$
and low $Q^2$ is necessary to describe the $F_2$ HERA data in the full range of
$Q^2$. This mechanism corresponds to saturation of gluon density. 
 
We summarize this part of our study by comparing the two versions of the
unintegrated gluon density, linear and non-linear, in Fig.~\ref{fig:ugluon}
(middle).  The linear gluon for $k^2>1\,\text{GeV}^2$ corresponds to the fit
described above.  For the values of $k^2$ below $1\,\text{GeV}^2$, similarly to
what we did for the non-linear case, we parametrize our gluon, this time by
$\phi(x,k^2)=\phi(x,k_0^2=1\,\text{GeV}^2)$.  We see in Fig.~\ref{fig:ugluon}
(middle) that at large values of $x$ and $k_t$, both distributions are similar.
As one goes to lower $k_t$, however, the linear gluon rises much faster than the
non-linear one. This effect becomes significantly stronger for smaller values of
$x$. 

\section{Central-forward dijet production in p-p collisions at the LHC}
\label{sec:results-pp}

\begin{figure}[t]
  \centering
  \includegraphics[width=0.45\textwidth,angle=-90]{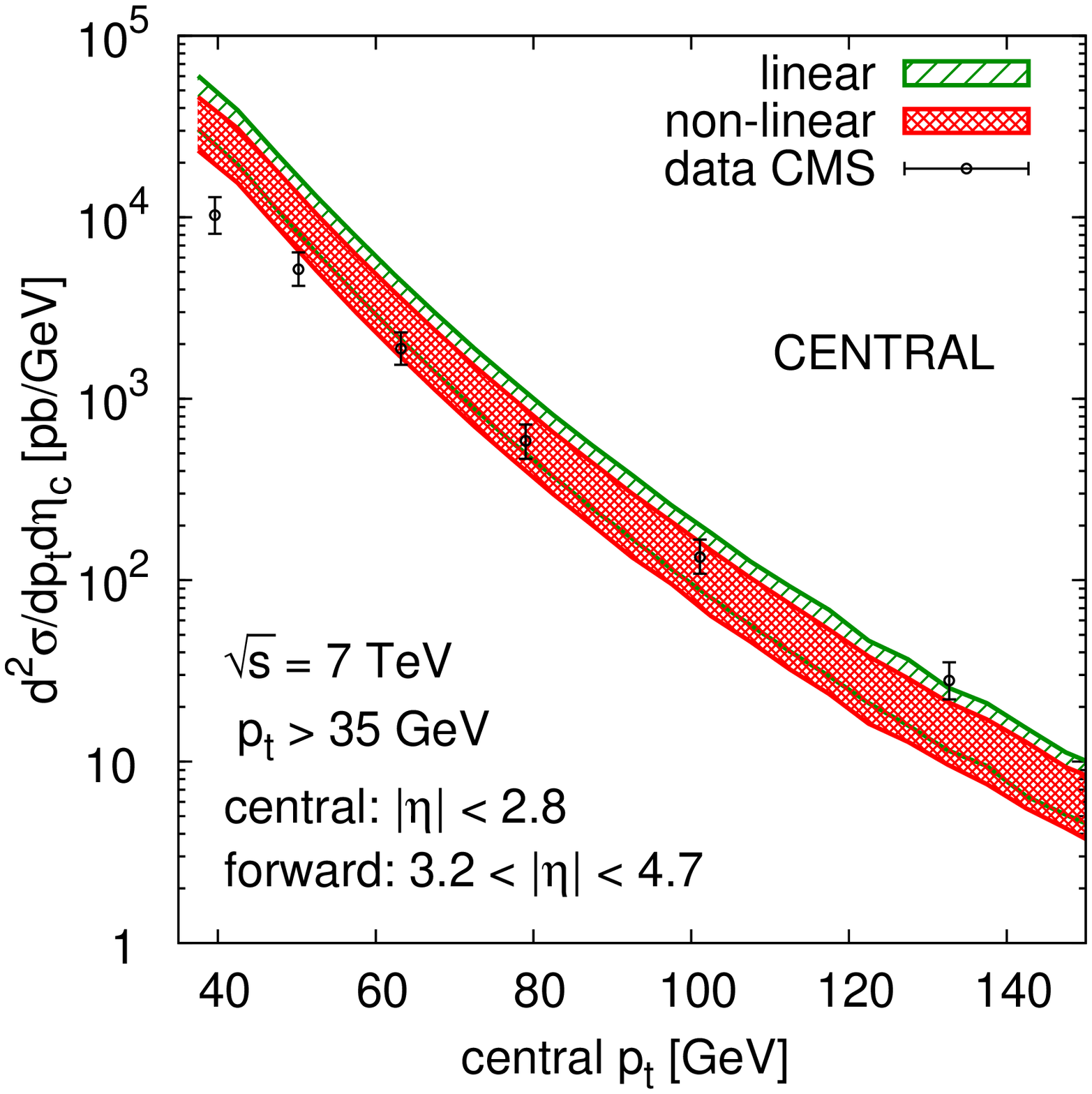}
  \hfill
  \includegraphics[width=0.45\textwidth,angle=-90]{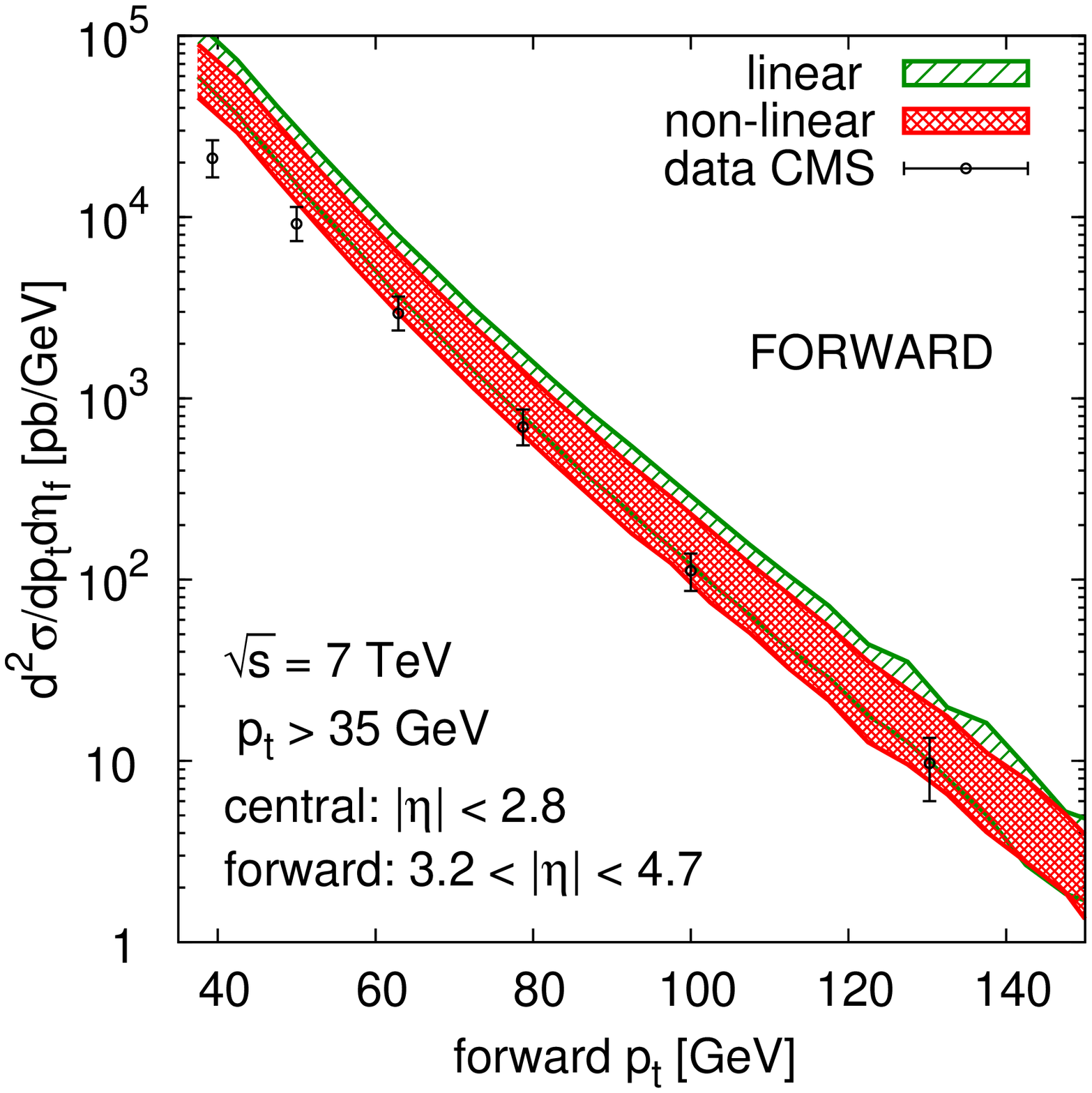}
  \caption{
  The transverse momentum spectra of the central (left) and forward (right) jets
  from our framework compared to the CMS data. The uncertainty bands come from
  varying the factorization and renormalization scale by factor two around the
  central value of $\mu = 60\, \text{GeV}$.
  }
  \label{fig:pt-spectra-cms}
\end{figure}

We are now in the position to compute the cross section for central-forward jet
production in the proton-proton collision. We use directly the formula
(\ref{eq:cs-main}) with the matrix element calculated in~\cite{Deak:2009xt} and
the unintegrated gluon distributions determined in the previous section. For the
collinear parton distributions we take CTEQ6mE~\cite{Pumplin:2002vw}.

The CMS collaboration has reported the measurement \cite{CMS-centfwd} of the
transverse momentum distributions of pairs of jets, one of which is restricted
to be in the forward and the other in the central part of the detector. 
This measurement has already been confronted with various Monte Carlo
predictions including HEJ~\cite{Andersen:2009nu, Andersen:2011hs} and
CASCADE~\cite{Jung:2010si}.
We compute the corresponding distributions in our framework taking the selection
cuts which match those of CMS.
We set the normalization and factorization scale equal to the fixed value of
$60\, \text{GeV}$ and vary it by the factors $1/2$ and $2$ to asses the
uncertainty.

The results for jet $p_t$ spectra are shown in Fig.~\ref{fig:pt-spectra-cms} for
the central~(left) and the forward~(right) jet respectively.  We find that our
predictions based on nonlinear (red) and linear (green) evolution equations
reproduce the pattern of the CMS measurement. As expected, this observable in
experimentally accessible region is weakly sensitive to saturation. We see some
suppression of the cross section based on the nonlinear evolution equation which
is a manifestation of the lower gluon density in the small $k_t$ region.
The description with both densities is good despite the fact that our modeling
of jets is very simple since, in the framework we use here, each of the two jets
is just a single parton.  Moreover, the small excess of low-$p_t$ is expected
precisely because of that. Adding a parton shower on the top of our partonic
result would cause some energy to go out of the jet and that would affect the
low-$p_t$ jets since they are broader then the high-$p_t$ jets.  For related
discussion within the HEJ approach see \cite{Andersen:2011hs,Andersen:2011zd}.
Similarly, we may expect that due to large rapidity gap between the produced
jet, some extra BFKL-type radiation could be emitted.
An explicit demonstration of the role of these effects in our framework opens an
interesting possibility for future work. However, they are not crucial for the
following discussion since they will affect the results for p-p and p-Pb in the
same way and our focus in this work is on the relative differences between
this two cases.
 
Also the slightly smaller cross section in the high-$p_t$ region for the central
jet is well understood. As follows from Eq.~(\ref{eq:x1x2}), the value of $x_2$
probed by the central jet $p_t \sim 140\, \text{GeV}$ corresponds to $x_2 \sim
0.02$ which is beyond the limit we used in our fits of unintegrated gluon and
therefore our predictive power in this region is limited. We have checked
that if the gluon is fitted with the upper limit on $x$ extended to
$0.02$, the result for $p_t$ distribution of the central jet becomes consistent
with the data also in the high-$p_t$ region. Trying to extend our framework to
higher values of $x$ would be in itself an interesting project. It goes however
beyond the scope of this paper, therefore throughout this study, we restrict
ourselves to the region of $x<0.01$. 
It should be emphasized, that even without going to large values of $x$
the distributions from Fig.~\ref{fig:pt-spectra-cms} are unique since they give
direct access to the unintegrated gluon at medium and large values of $k_t$,
provided that one works in the framework which permits to compute gluon in that
region. The framework we adopted for this study satisfies that criterion. That
opens the possibility to study the importance of terms of higher order from
the point of view of BFKL, i.e. energy conservation and subleading pieces of the
splitting function.

\section{Signatures of saturation in central-forward dijet production
in p-Pb collisions} 
\label{sec:results-p-Pb}

In the preceding sections, we have shown that the {\it high energy
factorization} formalism  with the unintegrated gluon from the QCD evolution
equation with saturation can successfully account for the features measured both
in the e-p and p-p collisions. The slightly less precise description of the data
in the latter case is expected and fully understood, and it can be traced
back to the purely partonic nature of our result or to the upper limit on the
$x$ used in our fits of unintegrated gluon.
 
One of the main features of the formalism we use is that the unintegrated gluon
is determined from the nonlinear evolution equation and therefore it exhibits
saturation effects around certain momentum scale $Q_s(x)$.
In this section, we address the question whether those effects could be studied
experimentally in collisions at the LHC. 

The main challenge of such a study, as pointed out by numerous analysis of HERA
data~\cite{GolecBiernat:1998js,Kowalski:2003hm,Bartels:2002cj,Iancu:2003ge,Albacete:2010sy},
is that the saturation scale in the proton seems to be of the order of a few
GeV, hence it lies just at the border between perturbative and non-perturbative
regime of QCD. This makes it difficult to access both theoretically and
experimentally.
One way to improve the situation is to go to the p-A collisions since, as widely
discussed in the literature, the saturation scale in the nucleus is expected to
be significantly higher than in the proton
\cite{Albacete:2010pg,Dominguez:2011cy,Stasto:2011ru,Tuchin:2009nf,Gelis:2010nm}.
To estimate the possible effects of saturation in the heavy nucleus we use a simple formula
for the  nucleus radius following from counting the number of nucleons for Woods-Saxon nuclear density
profile. The radius of the nucleus reads
\be
R_{\text{A}}=R\, A^{1/3} \,,
\label{eq:radius}
\ee 
where $R$ is the proton radius, which is one of the fitted parameters of our
framework as described in section~\ref{sec:fit} and $A$ is the mass number
($A=207$ for Pb, $A=196$ for Au).  The above definition has the property that in
the limit $A\rightarrow 1$ the result for the proton is recovered.

Analogous equation to Eq.~(\ref{eq:fkovres}) for the Heavy Ion (HI)  normalized
to the proton reads
\begin{multline} 
\phi_{\text{HI}/A}(x,k^2) \; = \; \phi_{\text{HI}/A}^{(0)}(x,k^2) \\
+\,\frac{\alpha_s(k^2) N_c}{\pi}\int_x^1 \frac{dz}{z} \int_{k_0^2}^{\infty}
\frac{dl^2}{l^2} \,   \bigg\{ \, \frac{l^2\phi_{\text{HI}/A}(\frac{x}{z},l^2) \, \theta(\frac{k^2}{z}-l^2)\,   -\,
k^2\phi_{\text{HI}/A}(\frac{x}{z},k^2)}{|l^2-k^2|}   +\,
\frac{k^2\phi_{\text{HI}/A}(\frac{x}{z},k^2)}{|4l^4+k^4|^{\frac{1}{2}}} \,
\bigg\} \\  + \, \frac{\alpha_s (k^2)}{2\pi k^2} \int_x^1 dz \,\Bigg[ 
\left(P_{gg}(z)-\frac{2N_c}{z}\right) \int^{k^2}_{k_0^2} d l^{
2}\,\phi_{\text{HI}/A}\left(\frac{x}{z},l^2\right)+zP_{gq}(z)\Sigma_{HI/A}\left(\frac{x}{z},k^2\right)\Bigg] \\
-\frac{2A^{1/3}\alpha_s^2(k^2)}{R^2}\left[\left(\int_{k^2}^{\infty}\frac{dl^2}{l^2}\phi_{\text{HI}/A}(x,l^2)\right)^2
+
\phi_{\text{HI}/A}(x,k^2)\int_{k^2}^{\infty}\frac{dl^2}{l^2}\ln\left(\frac{l^2}{k^2}\right)\phi_{\text{HI}/A}(x,l^2)\right]\,,
\label{eq:fkovres2} 
\end{multline} 
where we used Eq.~(\ref{eq:radius}) to express the radius of the heavy ion in
terms of the proton radius thus $\phi_{\text{HI}}(x,k^2)\equiv
A\phi_{\text{HI}/A}(x,k^2)$ is the distribution of gluons per nucleon in the
nucleus. 
We see that the strength of the nonlinear term in Eq.~(\ref{eq:fkovres2}) is
enhanced by $A^{1/3}$ \cite{JalilianMarian:2005jf,Gelis:2010nm}. We are aware
that this modification is not sufficient to fully model the nuclear target and
is a somewhat crude approximations but it will suffice as a first approximation
to estimate the saturation effects in the nucleus. For a more sophisticated
approach see \cite{Schenke:2012wb} and references therein.  

\begin{figure}[t]
  \centering
  \includegraphics[height=0.5\textwidth,angle=-90]{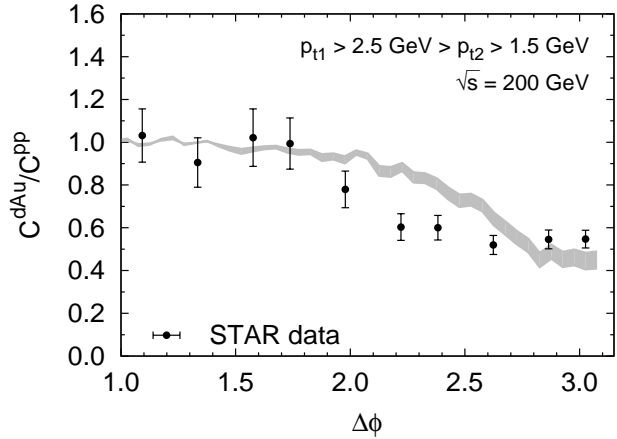}
  \caption{
  Ratio of d-Au/p-p coincidence probabilities $C(\Delta\phi)$ for the forward
dihadron production at RHIC as a function of the azimuthal distance between the
particles. The d-Au data were shifted by a constant. The errors of the ratio
were determined from relative errors of each $C(\Delta\phi)$ before the shift.
The band corresponds to our prediction with the uncertainty related the unknown
yield of uncorrelated dihadron production. 
}
  \label{fig:decor-d-Au}
\end{figure}

\begin{figure}[t]
  \centering
  \includegraphics[width=0.32\textwidth,angle=-90]{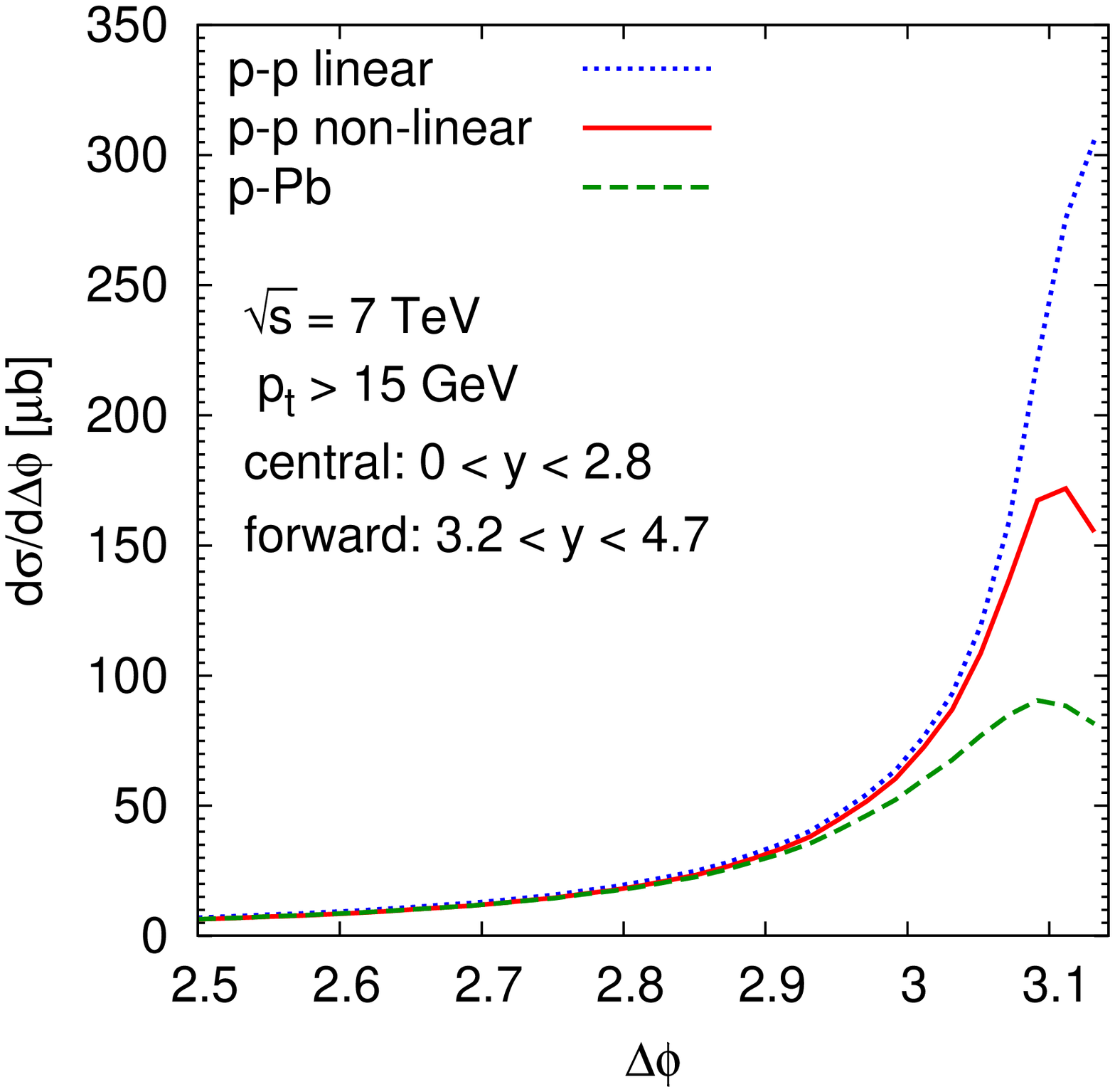}
  \includegraphics[width=0.32\textwidth,angle=-90]{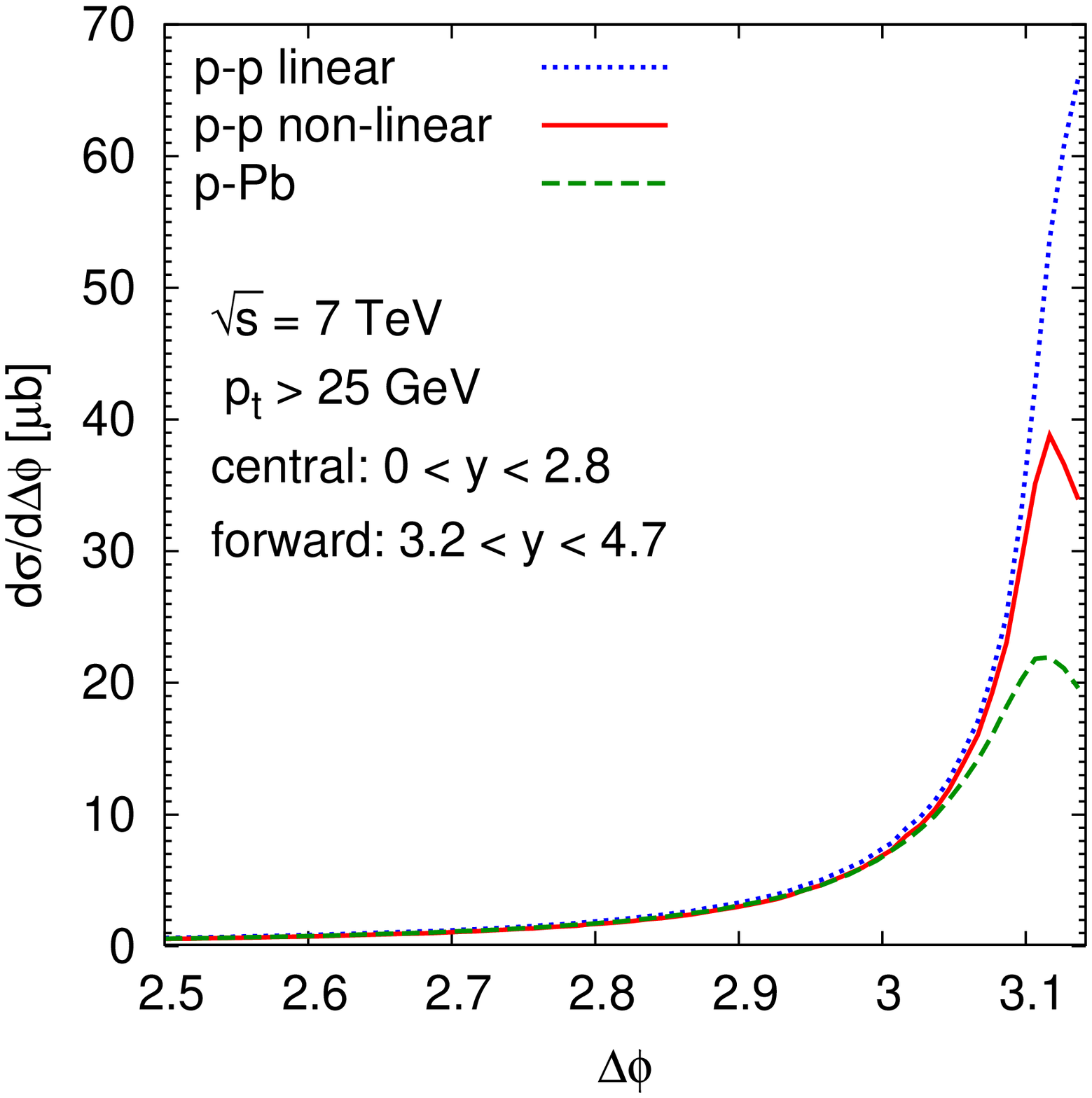}
  \includegraphics[width=0.32\textwidth,angle=-90]{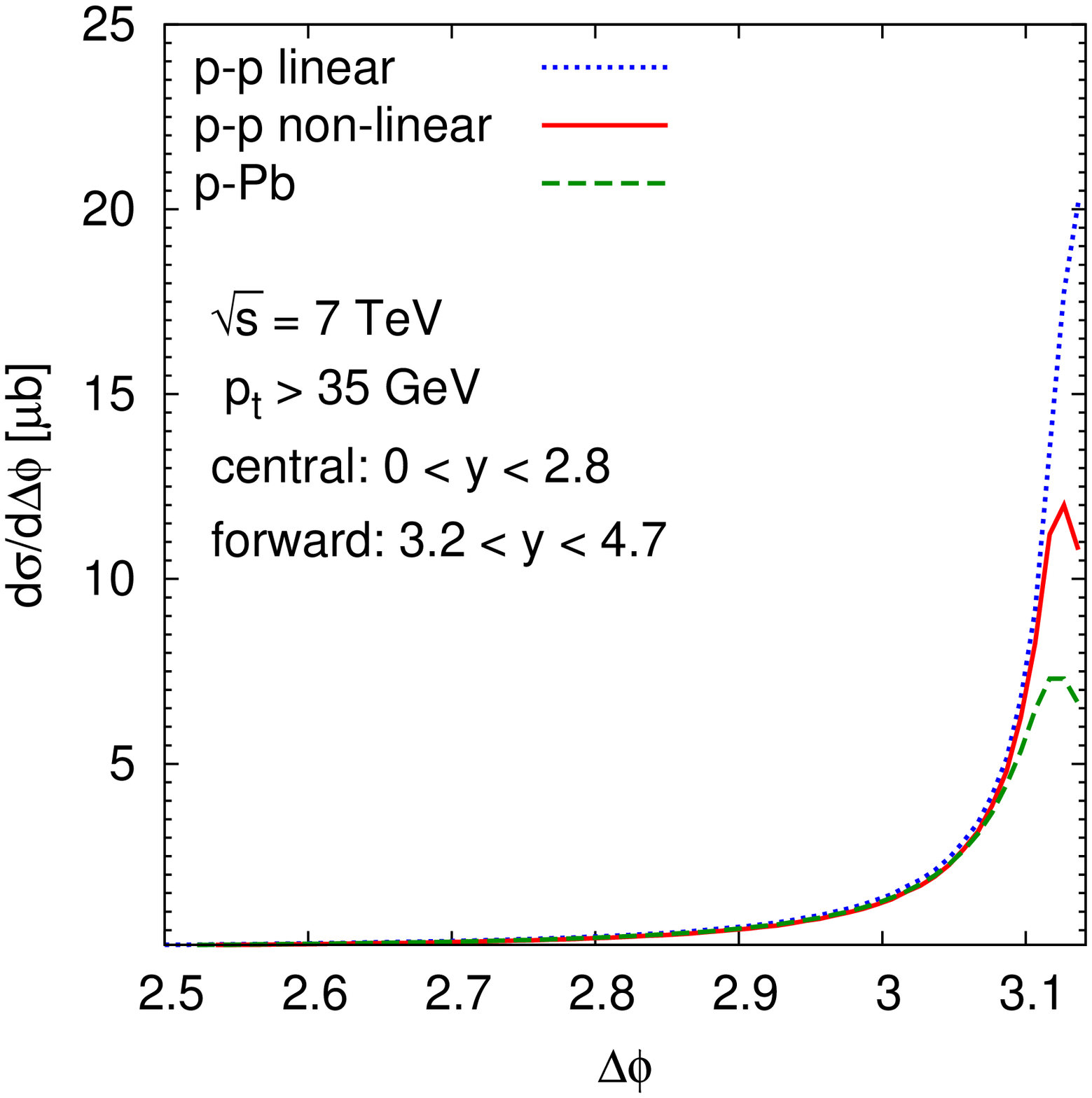}
  \includegraphics[width=0.31\textwidth,angle=-90]{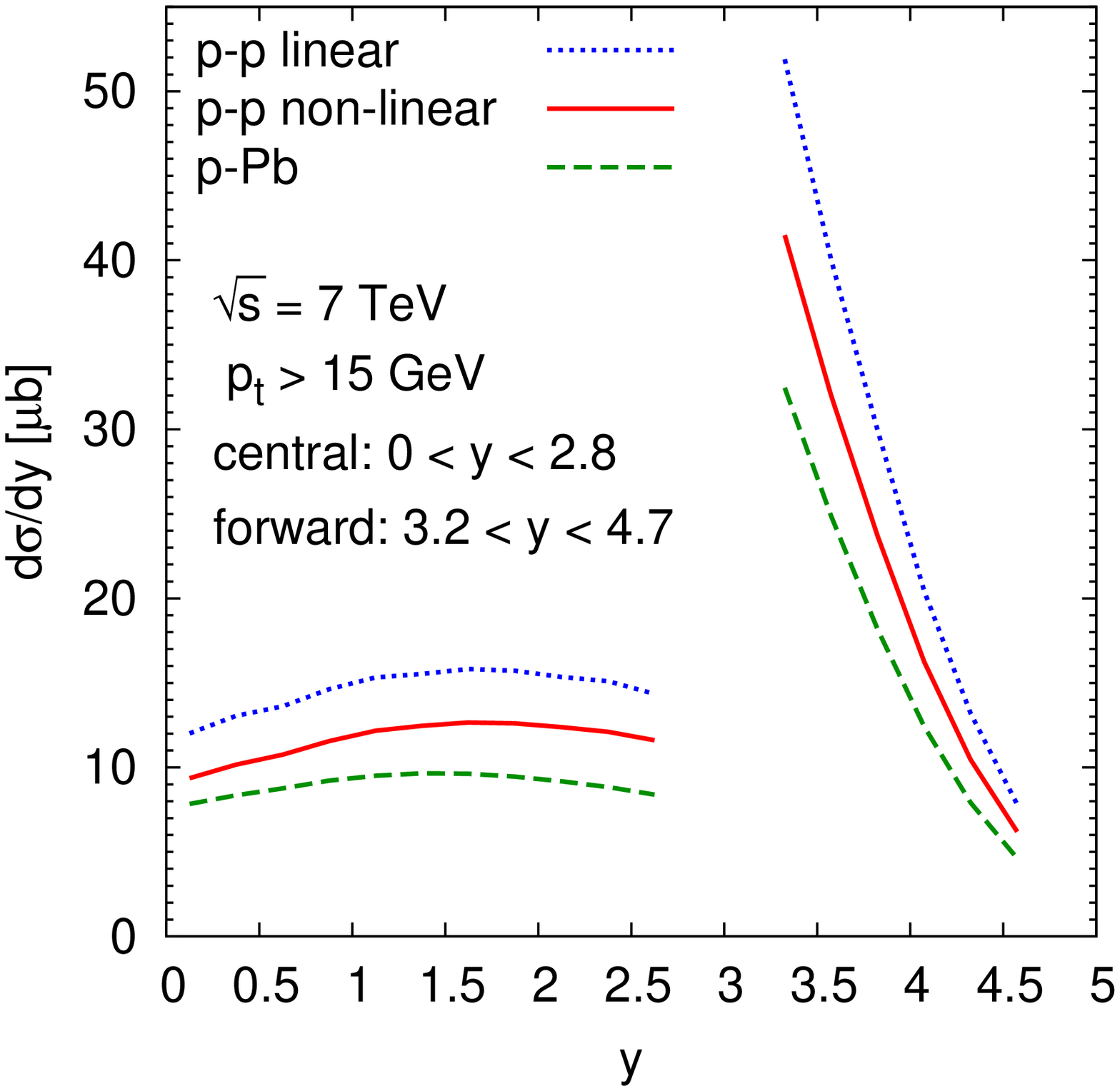}
  \includegraphics[width=0.31\textwidth,angle=-90]{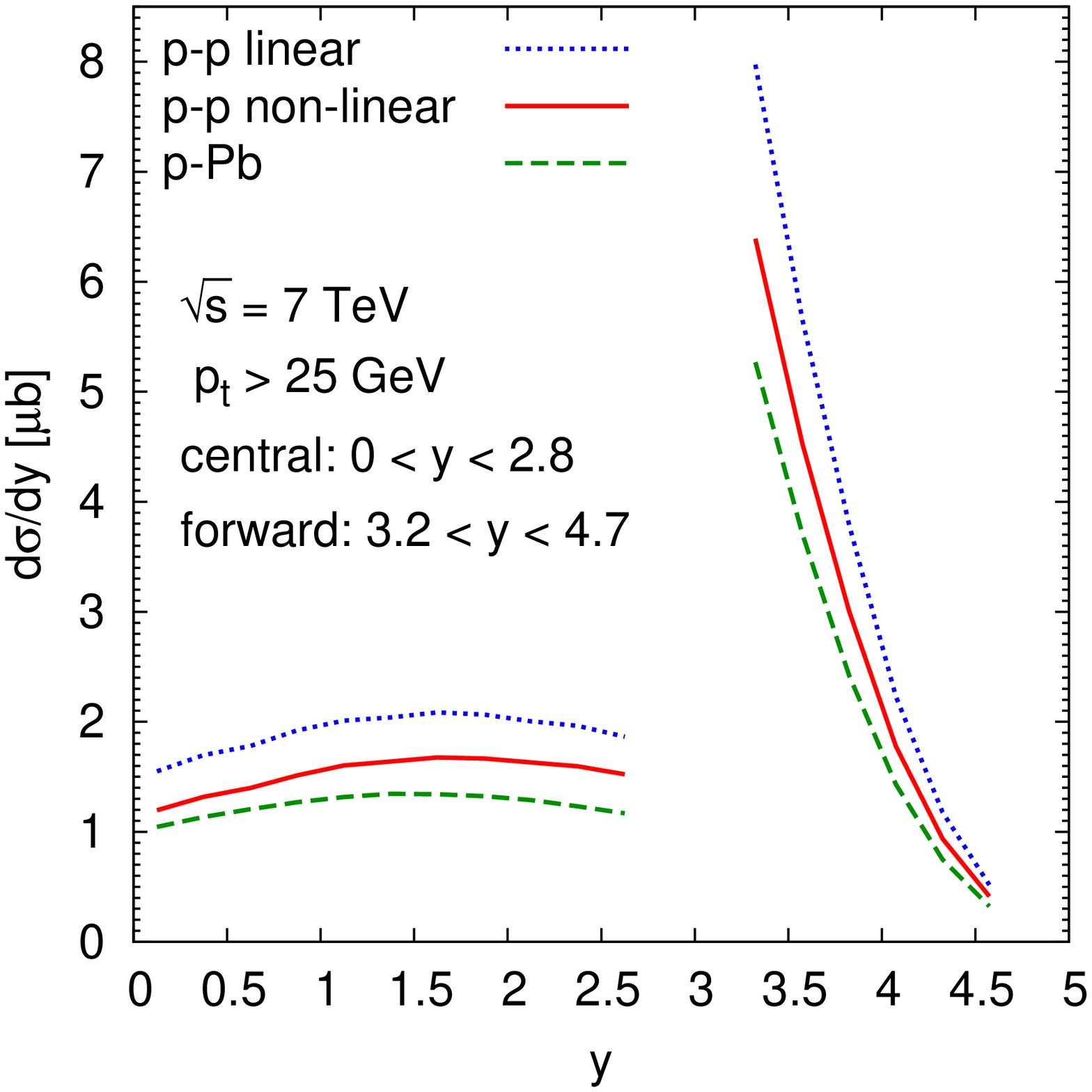}
  \includegraphics[width=0.31\textwidth,angle=-90]{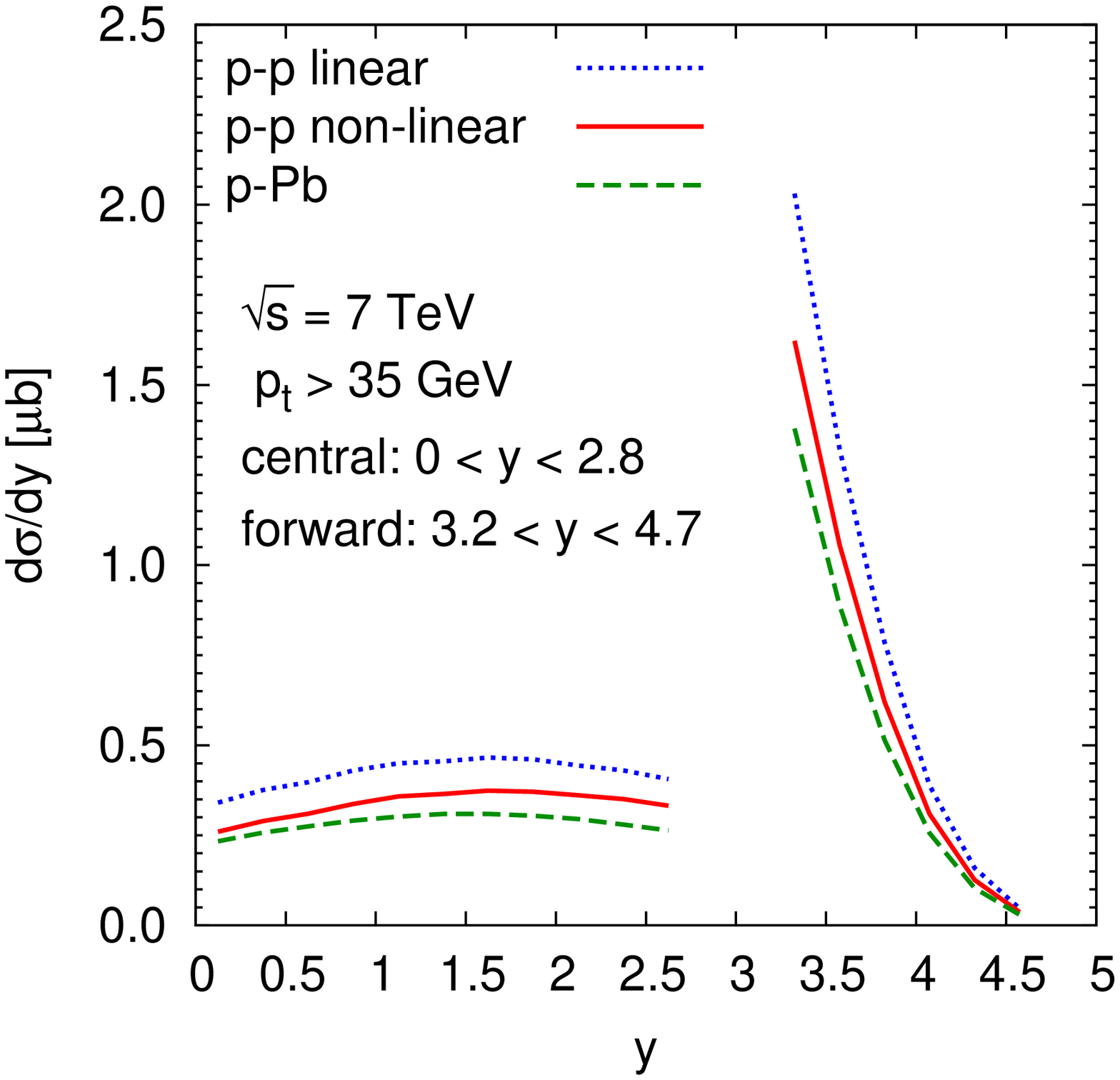}
  \caption{
  Differential cross sections for central-forward dijet production as
  functions of azimuthal distance between the jets $\Delta\phi$ (top) or
  rapidities of the jets (bottom) for the case of p-p and p-Pb collisions and 
  three different cuts on jets' $p_t$.
  \label{fig:decor-p-Pb}
  }
\end{figure}

\begin{figure}[t]
  \centering
  \includegraphics[width=0.32\textwidth,angle=-90]{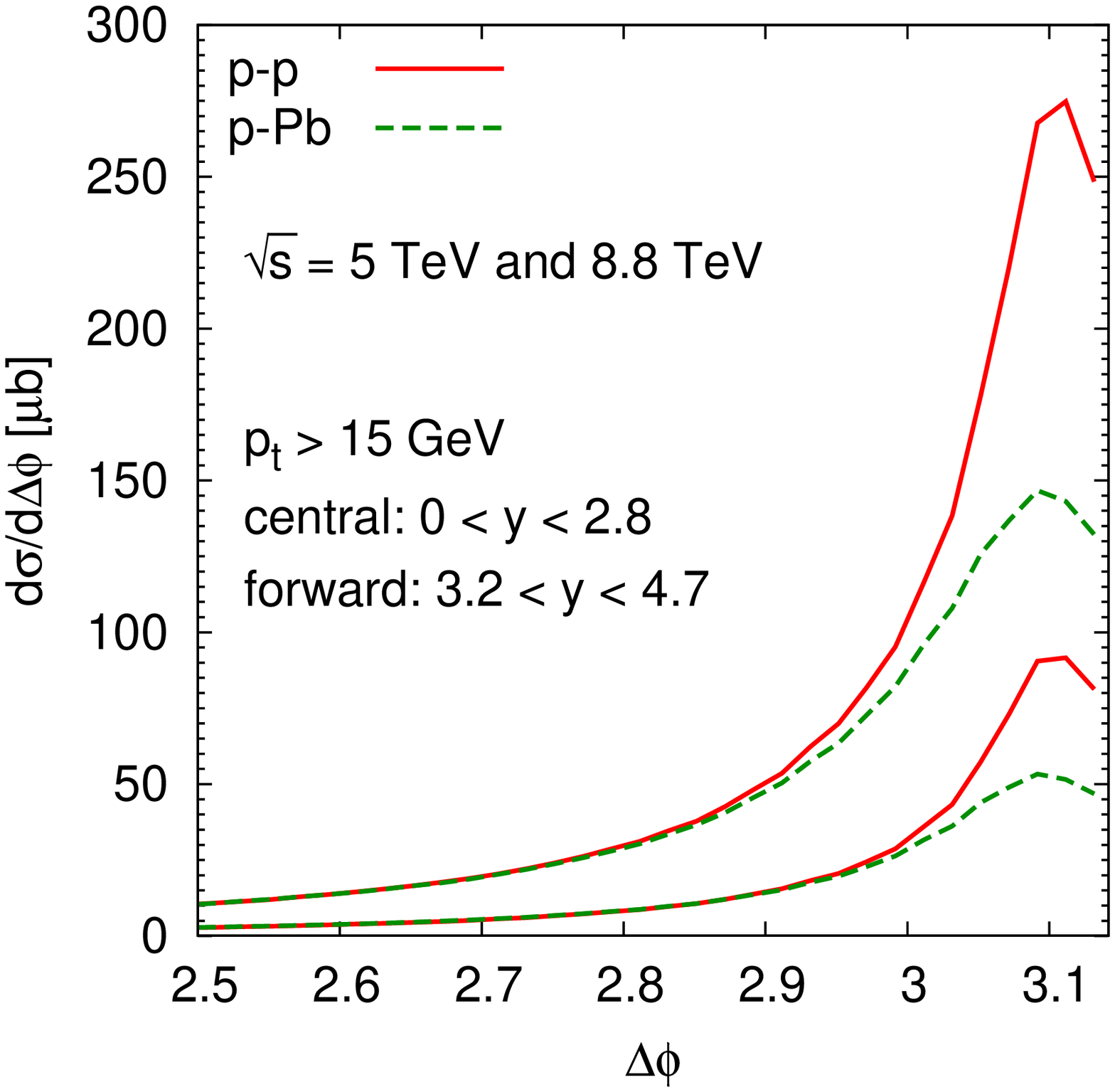}
  \includegraphics[width=0.32\textwidth,angle=-90]{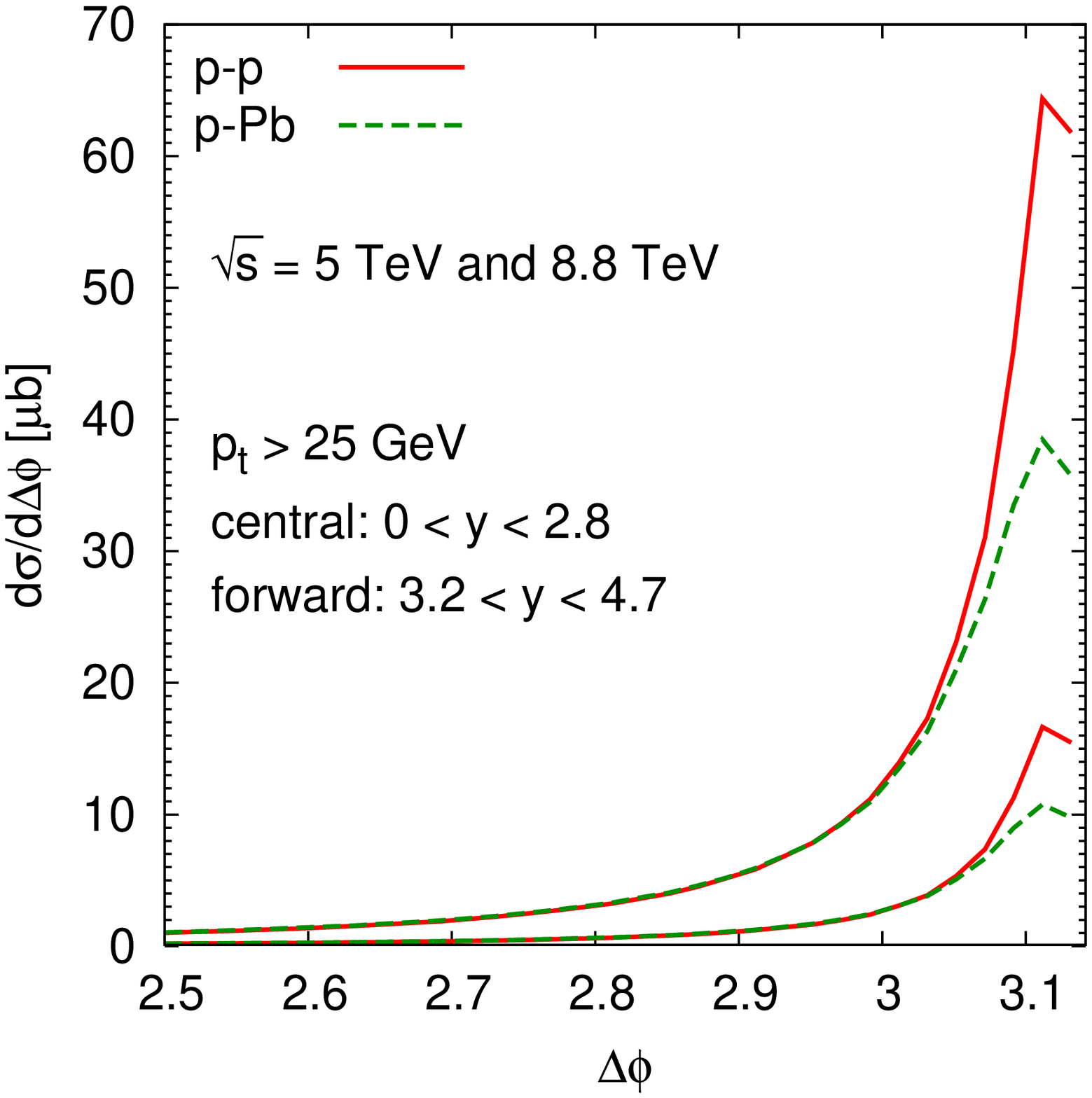}
  \includegraphics[width=0.32\textwidth,angle=-90]{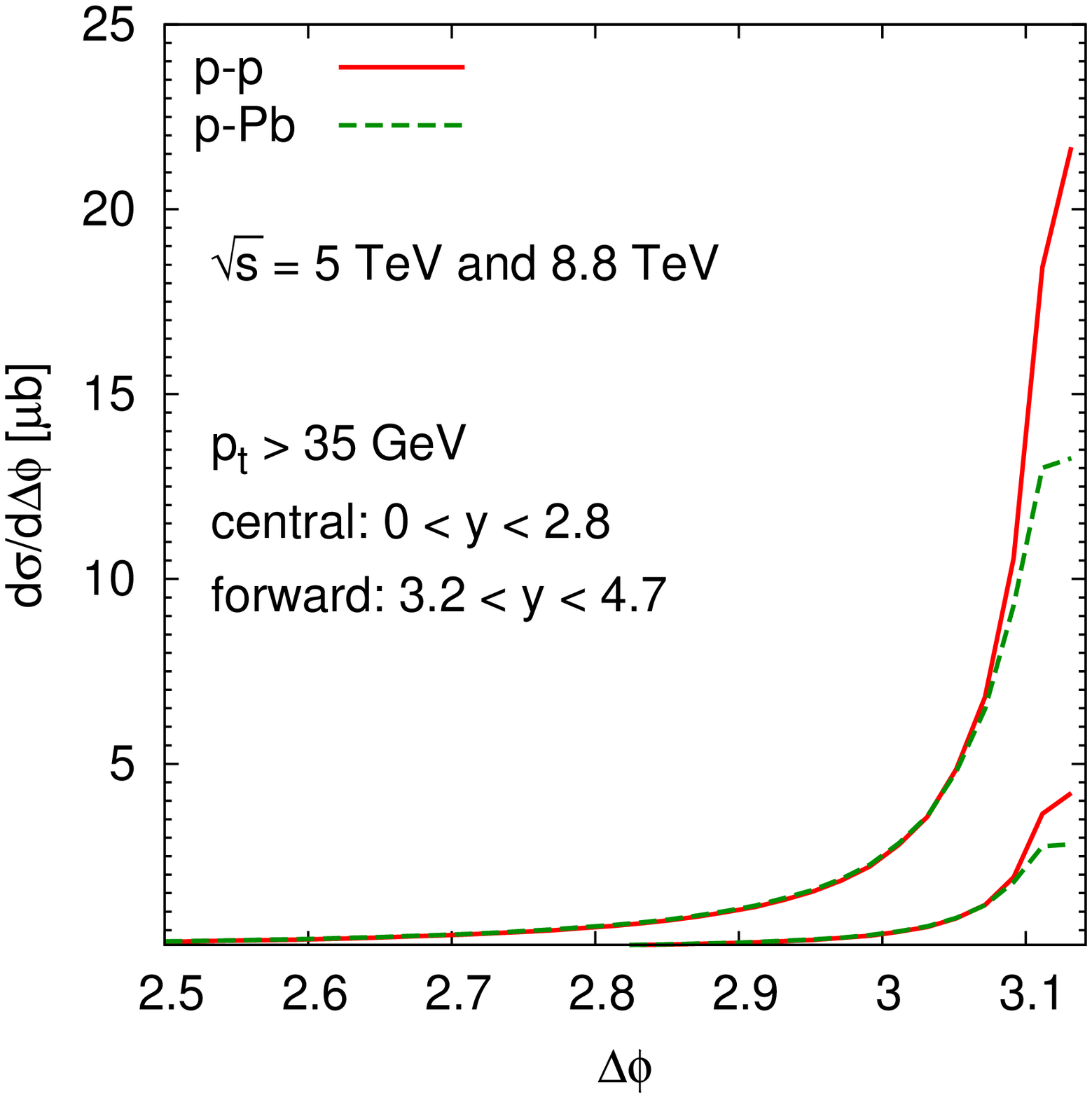}
  \caption{ 
  Differential cross sections for central-forward dijet production at $\sqrt{s}
= 5 \text{ TeV}$ and 8.8~TeV as functions of azimuthal distance between the jets
$\Delta\phi$ for the case of p-p and p-Pb collisions and three different cuts on
jets' $p_t$.
}
  \label{fig:decor-p-Pb-588}
\end{figure}

In Fig.~\ref{fig:ugluon} (right) we show the gluon density in the Pb nucleus 
which results from the application of the above prescription to the unintegrated
gluon distribution in the proton from Fig.~\ref{fig:ugluon} (left). We notice
that due to stronger saturation effects in Pb, the gluon density is lower than
that in the proton for the same value of $x$. We also see that in Pb the maxima
are shifted towards larger values of the gluon's transverse momentum~$k_t$,
which corresponds to the larger saturation scale in the nucleus.

An observable which is very well suited to study saturation is the azimuthal
correlation of the central and forward jet.  
It is an inclusive observable that measures radiation between jets and is
therefore sensitive to potential saturation effects which are supposed to
decrease the rate of emissions when the parton density is probed at very low $x$
where the high density of partons leads to their recombinations. 
The most interesting region of $\Delta\phi$ is that close to $\pi$ since
the produced jets are almost back to back and the gluon density is probed at 
low $k_t$. 

Before we turn to the discussion of our main results for the central-forward
dijets production at the LHC, it is interesting to check how the magnitude of
the suppression of decorrelations predicted by our framework compares to the
existing data for the forward dihadron production from RHIC. STAR and PHENIX
measured the coincidence probability defined as the ratio of the yield of the
$\pi^0$-pair to the inclusive $\pi^0$ yield, $C(\Delta\phi) =
N^\text{pair}(\Delta\phi)/N^\text{incl}$.
Precise determination of this observable requires convolution of our diparton
pair from Eq.~(\ref{eq:cs-main}) with the fragmentation functions as well as
computation of inclusive $\pi^0$ production and also the uncorrelated $\pi^0$
pair production (see \cite{Albacete:2010pg,Stasto:2011ru} for corresponding
results in different frameworks).  All this goes beyond the scope of the present
paper and will be published elsewhere. 
Nevertheless, we can make a meaningful estimate for a related quantity that can
be compared to the STAR data~\cite{Braidot:2010zh,Braidot:2011zj}. 
To minimise the contribution from the uncorrelated pair production we shift the
d+Au data by -0.00145, similarly to what has been done in
~\cite{Albacete:2010pg}, so that the d+Au and p-p data coincide below
$\Delta\phi=\pi/2$. Assuming that the remaining yield comes predominantly from
the correlated dihadron production, we compare the ratio
$C^\text{dAu}(\Delta\phi)/C^\text{pp}(\Delta\phi)$ for the experimental data
with that from our computation for diparton production at $\sqrt{s}=200\,
\text{GeV}$ with $p_{t1}>2.5\, \text{GeV}$, $p_{t1} >p_{t2}>1.5\, \text{GeV}$
and $2.4 < y_{1,2} < 4.0$. 
By using the above ratio we do not need to worry about the normalization to the
inclusive spectra and it is also reasonable to expect that most of the effects
from parton fragmentation will cancel.
Our absolute prediction (with $\mu=p_{t1}$) for the ratio of the coincidence
probabilities is shown in Fig.~\ref{fig:decor-d-Au} together with the data from
STAR~\cite{Braidot:2010zh,Braidot:2011zj}. 
Since our framework does not depend on the impact parameter, we chose to compare
to the data averaged over centralities. The band in Fig.~\ref{fig:decor-d-Au}
corresponds to the uncertainty related to the assumption of negligible
contribution form the uncorrelated production after the shift described above.
To asses this uncertainty we considered an alternative scenario in which half of
the yield seen in the data at $\Delta\phi=\pi/2$ is attributed to the
independent production and the other half to the correlated production that we
can predict with our framework. 
As we see in Fig.~\ref{fig:decor-d-Au}, the suppression pattern of the away-side
peak of the dihadron spectra from d-Au collisions at RHIC is correctly
reproduced by our calculation which shows that our theoretical framework
captures the essential physics of this class of processes.

We move now to the central-forward dijet production in the p-Pb collisions at
the LHC. In the top row of Fig.~\ref{fig:decor-p-Pb} we show the differential
cross section for the central-forward dijet production as a function of the
azimuthal distance between the jets for the p-p and p-Pb collisions. To obtain
those results we employed the linear and nonlinear versions of the evolution
equation~(\ref{eq:fkovres}) for the proton and the evolution equation~(\ref{eq:fkovres2}) for Pb. 
We used selection similar to that form the previous section except for the $p_t$
cut which we now vary from $15$, through $25$, to $35$ GeV and the rapidity
which is restricted to positive values. The latter corresponds to the fact that,
contrary to the p-p case, the p-Pb collision is asymmetric and, as follows from
Eq.~(\ref{eq:x1x2}), one probes the gluon density in Pb at low $x$ only by
measuring the forward jets going into the region of positive rapidity.
  
First observation from Fig.~\ref{fig:decor-p-Pb} is that the non-linear
evolution leads to a significant suppression of the $\Delta\phi$ and rapidity
distributions already for the proton case. This alone is a clear manifestation
of saturation.
Then we see that the $\Delta\phi$ cross section near the peak region is
suppressed further by the factor of about two for the case of the p-Pb collision
and the effect extends to lower values of $\Delta\phi$ as we lower the $p_t$
threshold (going from right to left plot). 
This is precisely the consequence of gluon saturation which is stronger in the
Pb nucleus then in the proton and therefore the unintegrated gluon distribution
in the region of small and medium $k_t$ is suppressed in Pb compared to the
proton case as shown in Fig.~\ref{fig:ugluon}~(right).  It is this region of
gluon's $k_t$ that is probed by the dijet configurations with $\Delta\phi \sim
\pi$ and that is what leads to the lower cross section in the area of the peak.

We notice that the non-linear results have a dip structure near $\Delta\phi
\simeq \pi$. This is a consequence of the feature of a high energy factorisable
gluon density which goes down to zero like $k^2$.  On the other hand, as
discussed in section~\ref{sec:fit}, in the linear case we model the behaviour of
the unintegrated gluon density by assuming that the gluon density behaves like a
constant and therefore the linear result for the $\Delta\phi$ distribution keeps
growing as $\Delta\phi\rightarrow\pi$.
These features make the $\Delta\phi$ distribution a particularly interesting
observable for testing shapes of gluon densities and more generally the validity
of the high energy factorization as pointed out recently in
\cite{Stasto:2012ru}.
On the top of that, the region near $\Delta\phi \simeq \pi$ is also sensitive to
Sudakov (virtual corrections) and parton shower effects (taking energy from the
jets we measure) which have the tendency to reduce the cross section in the
region $\Delta\phi$ near the peak. Some refinement along those lines could be
envisaged in the future.  These effects act however in a similar way for the
proton and for the heavy ion since they affect the hard scattering. 
Because we are interested in searching for saturation effects in the initial
state parton density, it is legitimate to neglect them in this study and focus
on the relative difference between the cases of p-p and p-Pb collisions.
The main point we would like to emphasize here is that
the suppression due to saturation predicted in Fig.~\ref{fig:decor-p-Pb} is both
strong and it extends for large enough range in $\Delta\phi$ to allow for
experimental discrimination between the linear and non-linear scenario, even if
the very small region near $\Delta\phi=\pi$ may profit from further refinements.

In the bottom row of Fig.~\ref{fig:decor-p-Pb} we present the rapidity
distributions of forward and central jets for the case of p-p and p-Pb
collisions. As expected, the saturation effects which are stronger in the nucleus
than in the proton lead to lower yields both for the central and forward jet
production in the p-Pb collision. Consistently with the decorrelation results,
also here, the difference between p-p and p-Pb becomes more pronounced as one
lowers the value of the jet $p_t$ cut.

Finally, in Fig.~\ref{fig:decor-p-Pb-588} we show decorrelations plots similar
to those from Fig.~\ref{fig:decor-p-Pb} but for the energies of the actual p-Pb
collisions, i.e. the current $\sqrt{s} = 5\, \text{TeV}$ and the nominal
$\sqrt{s} = 8.8\, \text{TeV}$. As expected, the total yields increase with
energy but the relative difference between the p-p and p-Pb case seems to remain
similar.

\section{Conclusion and outlook}
\label{sec:conclusions}

We presented the analysis of e-p, p-p and p-Pb collisions in the framework
of high energy factorisation with the unintegrated gluon density given by the
nonlinear QCD evolution equation. We have shown that this formalism can
successfully account for features measured in e-p and p-p data. For comparison, 
we also performed calculations within the linear evolution framework and
discussed differences between results from the two scenarios.

We then used the above non-linear framework to provide an estimate of the
effects of gluon saturation in the nuclei. 
We presented predictions for the azimuthal decorrelations as well as the
rapidity distributions for the p-p and p-Pb collisions.
Our main finding is that saturation in the Pb nucleus has a potential to manifest
itself as a factor two suppression of the central-forward jet decorrelation in the
region of the azimuthal distance between the jets $\Delta\phi \sim \pi$. The
effect becomes more pronounced with lower cuts on jets' transverse momenta.

The framework used in our study allows for a number of refinements that would
lead to a better description of data as well as for more accurate predictions.
We could, for example, extend our analysis by introducing non-trivial impact
parameter dependence of the unintegrated gluon density. That would allow us to
study saturation effects as a function of centrality of the p-Pb collision.

Another interesting possibility for future work is opened by an ongoing
discussion on breakdown of the high energy factorization and related issue
of multiple definitions of the unintegrated gluon density.
As advocated in~\cite{Dominguez:2010xd,Dominguez:2011wm}, this
generalized description provides a framework which is better theoretically
motivated. There are two reasons why the study in such a framework would be
interesting. First of all, because that framework was derived only with
simplified matrix elements, whereas in our study we have the exact ones. That
would allow one to investigate the relative importance of the corrections to the
high energy factorization formula from section~\ref{sec:hef} compared to the
case with  exact kinematics.
The second reason is that the consensus as to which gluon distribution should be
used to study dijet production has not been reached yet. The definition of the
unintegrated gluon that we used in this study follows directly from Feynman
diagrams and, as argued in~\cite{Avsar:2012hj}, is therefore the valid form of
unintegrated gluon density. 
Hence, we performed our study in the framework of the high energy
factorization and the results from sections~\ref{sec:results-pp} and
\ref{sec:results-p-Pb} can then be used in the
future as a benchmark for further studies within extended formalisms.

Finally, we note that the approach used in this study is unique as it allows one
to study hard final states in the framework with saturation.
This offers the possibility to constrain the unintegrated gluon density in the
large range of momentum available at the LHC. 
Our current limit of $x<0.01$ could be extended to higher values of $x$ by using
the gluon density from the CCFM equation with saturation~\cite{Kutak:2011fu}.
That would allow for a better description of the high $p_t$ tail of jet spectra.

\section*{Acknowledgments}

We would like to thank Krzysztof Golec-Biernat for thorough reading of this
manuscript and for numerous valuable suggestions.
We acknowledge useful correspondence with Jeppe R. Andersen regarding the dijet
spectra.
We acknowledge correspondence with Hannes Jung and Maciej Misiura regarding
the FSR effects on dijet spectra.
We appreciate valuable discussions and correspondence with David d'Enterria, Grzegorz Brona,
Michal Deak, Francesco Hautmann, Cyrille Marquet, Andreas van Hameren, Hannes Jung and Pierre van Mehelen, Bowen Xiao.
During this research KK was supported by the Foundation for Polish Science with
the grant Homing Plus/2010-2/6. SS acknowledges the Foundation for Polish
Science for support of his stay in Instytut Fizyki J\k{a}drowej PAN during
realization of this project.


\end{document}